\documentclass[prd,twocolumn,superscriptaddress,altaffilletter,showpacs]{revtex4}
\usepackage{eurosym}
\usepackage{amssymb,amsmath}
\usepackage{comment}
\usepackage{graphicx}

\begin{document}

\title{Kerr black hole parameters and its distance from the Earth in terms
of directly measurable quantities of accretion disk}
\author{Mehrab Momennia}
\email{momennia1988@gmail.com}
\affiliation{Instituto de F\'{\i}sica y Matem\'{a}ticas, Universidad Michoacana de San
Nicol\'as de Hidalgo,\\
Edificio C--3, Ciudad Universitaria, CP 58040, Morelia, Michoac\'{a}n,
Mexico.}
\date{\today }

\begin{abstract}
We extract elegant and concise analytic formulae for the mass and rotation
parameters of the Kerr black hole as well as its distance from the Earth
only in terms of directly measurable quantities of the accretion disk
revolving in the black hole spacetime background. To this end, we consider
massive geodesic particles circularly orbiting the Kerr black hole in the
equatorial plane and emitting frequency-shifted photons toward a distant
observer. We calculate the frequency shift and \emph{redshift rapidity} at
the detector location, and by solving an inverse problem, we express the
Kerr black hole parameters and its distance from a distant observer in terms
of a handful of observable elements, such as frequency shift, aperture angle
of the telescope, and \emph{redshift rapidity}, a newly introduced concept
in \cite{mMpBaHuNepjc2024}. The aperture angle of the telescope (angular
distance) characterizes the emitter position on the sky, and the \emph{%
redshift rapidity} is an observable relativistic invariant representing the
proper time evolution of the frequency shift. The relations presented in
this article allow us to disentangle mass, spin, and distance to the black
holes in the Kerr spacetime background and obtain these parameters
separately. Our analytic formulae are valid on the midline and close to the
line of sight, and they can be directly applied to supermassive black holes
hosted at the core of active galactic nuclei orbited by water vapor clouds
within their accretion disks. The generic exact relations are valid for an
arbitrary point of the emitter's orbit, and they can be employed in black
hole parameter estimation studies.

\vskip3mm

\end{abstract}

\maketitle

\section{Introduction}

Even though black holes were first introduced as mathematical solutions to
the general theory of relativity in 1915, recent observations of
gravitational waves \cite{GW} and electromagnetic waves \cite{EHTM87,EHTSgr}
proved their presence in nature. Even before these convincing pieces of
evidence, tracking the star trajectories close to the center of the Milky
Way galaxy for decades has already provided sufficient data supporting the
existence of a supermassive black hole located at the center of our galaxy 
\cite{Eckart,Ghez,Gillessen,Morris} (see \cite{Naoz,Will}\ for recent
results).

On the other hand, an independent general relativistic method for obtaining
the black hole parameters has been invented in \cite{hanPRD2015}\ and
further developed in \cite{pBaHmMuNprd2022,KdS,mMpBaHuNepjc2024} with the
aim of utilizing observational frequency shift and \emph{redshift rapidity}
detected from massive test particles orbiting black holes. These types of
data differ from the ones employed in \cite{GW,EHTM87,EHTSgr}.
The initial approach \cite{hanPRD2015} has been introduced based on
kinematic redshift/blueshift in order to generalize the similar Keplerian
models \cite{Herrnstein2005,Argon2007,Humphreys2013}\ to general relativity.

Lately, by further extending this method, the mass and rotation parameters
of the Kerr black hole have been analytically expressed in terms of
observational redshift and blueshift as well as the radius of the emitter
that is not an observable\ quantity \cite{pBaHmMuNprd2022}. In addition, it
was shown that by taking into account the rotating Kerr black hole in a
spacetime background with de Sitter asymptote, it is possible to
obtain the Hubble (cosmological) constant and black hole parameters with the
help of this general relativistic method \cite{KdS}. Besides, the
polymerized black holes have been analyzed and their free parameters were
found in terms of the observational frequency shift \cite{FuZhang}.
Moreover, this method has been developed to generic static and spherically
symmetric black hole spacetimes in \cite{dMmMaHepjc2024}. This extension
allows expressing the free parameters of static black hole spacetimes in modified
gravities and general relativity theory coupled to various matter fields in
terms of observational frequency shift.

From an observational point of view, there are publicly available
observational frequency shift data from supermassive black holes hosted at
the core of the active galactic nuclei (AGNs) (see \cite%
{Herrnstein2005,Argon2007,Humphreys2013} for instance). In these
astrophysical systems, water vapor clouds are circularly orbiting the
central black hole within the accretion disk and water molecules emit
frequency-shifted photons toward a far away detector. In these specific
systems, the total frequency shift and aperture angle of the telescope are the observational quantities, and therefore,
one can employ this general relativistic approach to estimate the
Mass-to-Distance ($M/D$) ratio of the central black hole. Following this
capability, the mass-to-distance ratio of several supermassive Schwarzschild
black holes hosted at the core of AGNs has been estimated in \cite%
{ApJL,TXS,TenAGNs,FiveAGNs,TwoAGNs}.

It is worthwhile to mention that although the gravitational redshift
produced by the central compact object has been quantified in the $M/D$
estimations performed in \cite{ApJL,TXS,TenAGNs,FiveAGNs,TwoAGNs}, the
mass-to-distance ratio of the central supermassive Schwarzschild black hole
has been estimated only. This is because the mass-to-distance ratio is
degenerate in this general relativistic formalism and one has to use another
observational element to break the $M/D$\ degeneracy. In order to resolve
this issue, we have introduced a new general relativistic invariant
observable parameter called the \emph{redshift rapidity} that describes the evolution of the frequency shift with respect to the proper time \cite%
{mMpBaHuNepjc2024}. As a result, the $M/D$\ degeneracy in the Schwarzschild
spacetime background has been broken with the help of the \emph{redshift
rapidity}, and consequently, the mass of the Schwarzschild black hole and
its distance from the Earth have been expressed fully in terms of directly
observational quantities through concise and elegant analytic formulae.

A similar degeneracy is also present for the mass-to-distance ratio and
charge-to-distance ratio in the Reissner-Nordstr\"{o}m black hole spacetime.
Therefore, the \emph{redshift rapidity} has been computed in the
Reissner-Nordstr\"{o}m background and it has been used to break these
degeneracies in order to extract\ analytic formulae for the mass of the
Reissner-Nordstr\"{o}m black hole, its electric charge, and its distance
from a distant observer in terms of directly observational elements \cite%
{PabloRNBH}. The obtained relations for the Schwarzschild and
Reissner-Nordstr\"{o}m black hole solutions could help to measure the
distance to these black holes and compute their parameters. The initial aim
of introducing the redshift rapidity was to improve the
precision in measuring cosmic distances with respect to the previous
attempts based on post-Newtonian methods that measure the angular-diameter
distance to galaxies \cite{Humphreys2013,MCP2,Reid2019,MCPXI}. In this
study, we compute the \emph{redshift rapidity} in the Kerr background, and
with the aid of the total frequency shifts, we obtain concise and elegant
analytic formulae for the black hole mass, its spin, and its distance from a
distant observer fully in terms of directly observational quantities. The
complete expressions of the frequency shift and \emph{redshift rapidity} as
well as the analytic formulae could enable one to obtain the Kerr black hole
mass, spin, and its distance from the Earth.

The outline of this paper is as follows. The first part of the next section
is devoted to a review of the general relativistic method and the frequency
shift of massive geodesic particles circularly orbiting the Kerr black hole
in the equatorial plane. Then, we express the mass-to-distance ratio and the
spin-to-distance ratio of the Kerr black hole in terms of observable
frequency shifts on the midline and close to the line of sight (LOS). In
Sec. \ref{SecRedshiftRapidity}, we define \emph{redshift rapidity} in the
Kerr background as the derivative of the frequency shift with respect to the
proper time. Then, we obtain the \emph{redshift rapidity} on the midline and
close to the LOS, and with the aid of the corresponding mass-to-distance
ratio and spin-to-distance ratio, we\ express the mass and spin of the black
hole and its distance from the Earth in terms of direct observational
quantities. Finally, we finish our paper with some concluding remarks in
Sec. \ref{Discussion}.

\section{Frequency shift in Kerr background}

\label{Sec:KerrRedshift}

In this section, we first briefly review previous results from a general
relativistic method to obtain the frequency shift formulae of massive probe
particles revolving in Kerr spacetime background based on \cite%
{pBaHmMuNprd2022}, and then describe our original contribution. We first
consider the geodesic motion of massive test particles orbiting a Kerr black
hole which emit photons toward a distant observer. We present a formula for
the redshift of the detected photons coming from a general point of their
orbit in the equatorial plane which depends on the azimuthal angle $\varphi $%
. Then, we obtain the mass-to-distance ratio and spin-to-distance ratio of
the Kerr black hole in terms of directly measurable elements for two special
points on the midline $\varphi \approx \pi /2$ and close to the LOS $\varphi
\approx 0$.

The spacetime background of the Kerr black hole is described by the line
element 
\begin{equation}
ds^{2}=g_{tt}dt^{2}+2g_{t\varphi }dtd\varphi +g_{\varphi \varphi }d\varphi
^{2}+g_{rr}dr^{2}+g_{\theta \theta }d\theta ^{2},  \label{metric}
\end{equation}%
with the metric components 
\begin{equation}
g_{tt}=-\left( 1-\frac{2Mr}{\Sigma }\right) ,\quad g_{t\varphi }=-\frac{%
2Mar\sin ^{2}\theta }{\Sigma },\quad g_{rr}=\frac{\Sigma }{\Delta },  \notag
\end{equation}%
\begin{equation}
g_{\varphi \varphi }=\left( r^{2}+a^{2}+\frac{2Ma^{2}r\sin ^{2}\theta }{%
\Sigma }\right) \sin ^{2}\theta \,,\quad g_{\theta \theta }=\Sigma \,, 
\notag
\end{equation}%
\begin{equation}
\Delta =r^{2}+a^{2}-2Mr\,,\quad \quad \Sigma =r^{2}+a^{2}\cos ^{2}\theta \,,
\notag
\end{equation}%
where $M$ is the total mass of the Kerr black hole and $a$ is its total
angular momentum per unit mass, $a=J/M$ ($0\leq a\leq M$). The Kerr
spacetime has an intrinsic ring singularity of radius $a$ in the equatorial
plane, whereas its Cauchy horizon $r_-$ and event horizon $r_+$ surfaces are
located at%
\begin{equation}
r_{\pm }=M\pm \sqrt{M^{2}-a^{2}}.
\end{equation}

In axially symmetric spacetimes of the form (\ref{metric}), the frequency
shift of photons emitted by massive geodesic particles orbiting the black
hole and detected by an observer is given by \cite{hanPRD2015} 
\begin{eqnarray}
1 &+&z_{_{Kerr}}\!=\frac{\omega _{e}}{\omega _{d}}  \notag \\
&=&\frac{(E_{\gamma }U^{t}-L_{\gamma }U^{\varphi
}-g_{rr}U^{r}k^{r}-g_{\theta \theta }U^{\theta }k^{\theta })\mid _{e}}{%
(E_{\gamma }U^{t}-L_{\gamma }U^{\varphi }-g_{rr}U^{r}k^{r}-g_{\theta \theta
}U^{\theta }k^{\theta })\mid _{d}}\,,  \label{GenericShift}
\end{eqnarray}%
where $\omega _{p}=-\left( k_{\mu }U^{\mu }\right) \mid _{p}$\ is the
frequency of a photon emitted (detected) by an emitter (detector) at the
emission (detection) point $p$, $k_{p}^{\mu }$\ is the $4$-wave vector of
the photon, and $U_{p}^{\mu }$ is the proper $4$-velocity of the emitter
(observer). Besides, $E_{\gamma }$ and $L_{\gamma }$ are conserved along the
light trajectories and correspond to the total energy and axial angular
momentum of the photons, respectively. The relation (\ref{GenericShift}) holds for the frequency
shift of arbitrary stable orbits of the test massive particles, such as
circular, elliptic, non-equatorial, etc.

Now, because of the existing data regarding the accretion disks as well as
being able to extract analytic formulae for mass, spin, and distance to the
Kerr black hole, we restrict ourselves to the circular motion of the photon
sources ($U_{e}^{r}=0$) which is the case for real astrophysical systems
containing the accretion disks \cite{EdgeOn2011,ApJL}. In addition, we put
both the emitter and observer in the equatorial plane ($\theta =\pi /2$),
for which we have $U_{e}^{\theta }=0=U_{d}^{\theta }$, due to the fact that
any tilted disk should be driven to the equatorial plane of the rotating
spacetime backgrounds \cite{EquatorialPlane} ($U_{e}^{\theta }=0$) and
accretion disks can be detected mostly in an edge-on view from the Earth 
\cite{EdgeOn2011,EdgeOn2017}\ ($U_{d}^{\theta }=0$). In the special case of
a distant observer $r_{d}\rightarrow \infty $ (which is the case for the
real astrophysical systems in AGNs), the $4$-velocity of the detector
simplifies to $U_{d}^{\mu }=\delta _{t}^{\mu }$. By taking into account these assumptions, the generic
relation (\ref{GenericShift})\ reduces to \ 
\begin{equation}
1+z_{Kerr}\!=\frac{\left. \left( E_{\gamma }U^{t}-L_{\gamma }U^{\varphi
}\right) \right\vert _{e}}{\left. \left( E_{\gamma }U^{t}\right) \right\vert
_{d}}=U_{e}^{t}-b_{\varphi }\,U_{e}^{\varphi }\,,  \label{zcircorbits}
\end{equation}%
where $b_{\varphi }\equiv L_{\gamma }/E_{\gamma }$ is the deflection of
light parameter which gives the light bending produced by the gravitational
field of the Kerr black hole. The nonvanishing components of the $4$%
-velocity $U_{e}^{\mu }$ are given by \cite{pBaHmMuNprd2022} 
\begin{equation}
U_{e}^{t}(r_{e},\pi /2)=\frac{r_{e}^{3/2}\pm aM^{1/2}}{r_{e}^{3/4}\sqrt{%
r_{e}^{3/2}-3Mr_{e}^{1/2}\pm 2aM^{1/2}}},  \label{tCompVelo}
\end{equation}%
\begin{equation}
U_{e}^{\varphi }(r_{e},\pi /2)=\pm \frac{M^{1/2}}{r_{e}^{3/4}\sqrt{%
r_{e}^{3/2}-3Mr_{e}^{1/2}\pm 2aM^{1/2}}},  \label{phiCompVelo}
\end{equation}%
where $r_{e}$\ is the radius of the emitter, the upper sign refers to a
co-rotating emitter (the angular momentum of the massive test particle is in
the same direction as the Kerr black hole), and the lower sign corresponds
to a counter-rotating one. Also, one can calculate the $(\varphi +\delta )$%
-dependent light bending parameter $b_{\varphi }$\ for an arbitrary point of
the circular orbit of photon sources on the equatorial plane which is
presented in Eq. (\ref{bGamma}) of the appendix, and for the Kerr black hole
case, converts to [see also Fig. \ref{PhiDeltaFig} for the geometrical
illustration of the azimuthal angle $\varphi $ and the angular distance $%
\delta $, and their relation] 
\begin{eqnarray}
b_{\varphi } &=&-\frac{2aM}{r_{e}-2M}+\frac{r_{e}\Delta _{e}^{3/2}\sin
(\varphi +\delta )}{\left( r_{e}-2M\right) }\times   \notag \\
&&\frac{1}{\sqrt{\Delta _{e}^{2}\sin ^{2}(\varphi +\delta )+\left(
r_{e}-2M\right) r_{e}^{3}\cos ^{2}(\varphi +\delta )}},  \label{bVsPhiDelta}
\end{eqnarray}%
where $\Delta _{e}=\left. \Delta \right\vert _{r=r_{e}}$, $\varphi $ is the
azimuthal angle that is not a measurable quantity, and $\delta $ is the
aperture angle of the telescope (angular distance) that is an observable
parameter. It is worth mentioning that because of the dragging effect
produced by the rotation nature of the Kerr black hole, $b_{\varphi }$ does
not vanish at the line of sight where $\varphi =0=\delta $, in contrast to
the Schwarzschild case. In addition, the observable $\delta $\ is a function
of the azimuthal angle $\varphi $ for an arbitrary point on the circular
orbit through Eq. (\ref{deltaPhiRel})\ (see Fig. \ref{PhiDeltaFig} of the
appendix\ and related discussion).

Now, by substituting (\ref{tCompVelo})-(\ref{bVsPhiDelta}) into (\ref%
{zcircorbits}), we obtain the following explicit form of the frequency shift
for an arbitrary point of the orbit on the equatorial plane
\begin{widetext}
\begin{eqnarray}
&&1+z_{Kerr}\!\!=\!\frac{1}{\left( 1\!-\!2\tilde{M}\right) \sqrt{1\!-\!3%
\tilde{M}\pm 2\,\tilde{a}\,\tilde{M}^{\frac{1}{2}}}}\left[ 1\!-\!2\tilde{M}\!\pm \!\tilde{M}^{\frac{1}{2}}\left( \!\tilde{a}+    \frac{\tilde{\Delta}_{e}^{3/2}\sin (\varphi +\delta )}{\sqrt{%
\tilde{\Delta}_{e}^{2}\sin ^{2}\!(\varphi +\delta )+\!(1\!-\!2\tilde{M})\cos
^{2}\!(\varphi +\delta )}}\right) \right] ,  \label{zphi}
\end{eqnarray}
\end{widetext}
with $\tilde{M}=M/r_{e}$, $\ \tilde{a}=a/r_{e}$, and $\tilde{\Delta}%
_{e}=\Delta _{e}/r_{e}^{2}=1+\tilde{a}^{2}-2\tilde{M}$. In this relation,
the upper sign corresponds to the co-rotating photon courses and the lower
sign corresponds to the counter-rotating ones. In what follows we shall
restrict ourselves to the co-rotating emitters since the case of
counter-rotating ones is quite similar and one only needs following the same
steps as we discuss here for the co-rotating particles.

Besides, the innermost stable circular orbit (ISCO) radius in the Kerr
spacetime reads \cite{Bardeen} 
\begin{equation}
r_{ISCO}=M\left( 3+\beta \mp \sqrt{(3-\alpha )(3+\alpha +2\beta )}\right) ,
\label{rms}
\end{equation}%
\begin{equation*}
\alpha =1+\left( 1-\frac{a^{2}}{M^{2}}\right) ^{1/3}\left[ \left( 1+\frac{a}{%
M}\right) ^{1/3}+\left( 1-\frac{a}{M}\right) ^{1/3}\right] ,
\end{equation*}%
\begin{equation*}
\beta =\sqrt{\alpha ^{2}+3\frac{a^{2}}{M^{2}}},
\end{equation*}%
that approximately characterizes the inner edge of the accretion disk. In
this article, we are interested in stable circular orbits of the photon
sources such that $r_{e}\geq r_{ISCO}$,\ and this restriction on the emitter
radius leads to an upper bound on the frequency shift of the emitted photons
(see the dashed curves in Fig. 2 of \cite{pBaHmMuNprd2022}).

\begin{figure*}[t]
\centering
\includegraphics[scale=.95]{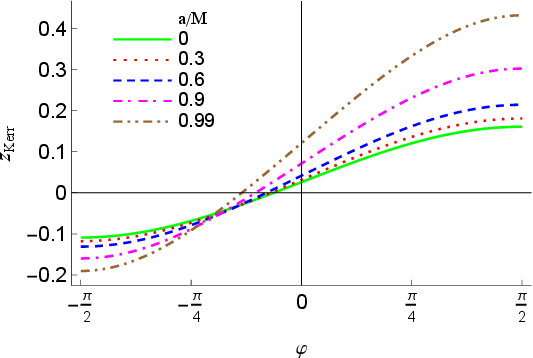} \includegraphics[scale=.95]{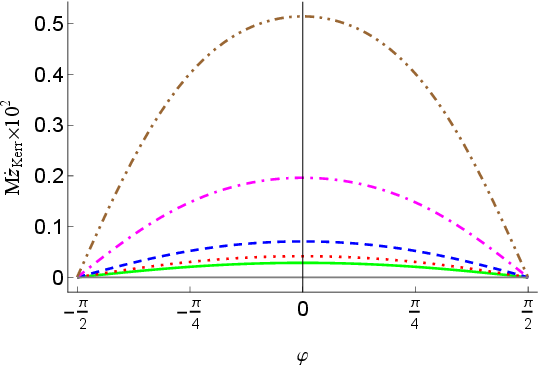}
\caption{The frequency shift $z_{Kerr}$ (left panel) and the \emph{redshift
rapidity} $\dot{z}_{Kerr}$ (right panel) versus the azimuthal angle $\protect%
\varphi$ in the Kerr black hole spacetime for $r_{e}=10 r_{ISCO}$, $D=10^{4}
r_{ISCO}$, and different values of the rotation parameter. The continuous green curves correspond to the Schwarzschild black hole, and the difference between the other curves and the green one is due to the dragging effect. The redshift and blueshift are maximal on the midline where $\protect\varphi \approx \pm 
\protect\pi /2$ while we see that the \emph{redshift rapidity} is maximum close to
the line of sight where $\protect\varphi \approx 0 \approx \protect\delta$.
In order to plot these curves, we substituted $\protect\delta $ given in Eq.
(\protect\ref{deltaPhiRel}) into the frequency shift formula (\protect\ref%
{zphi}) and the \emph{redshift rapidity} formula (\protect\ref{RedRap}).}
\label{zDotFig}
\end{figure*}

Figure \ref{zDotFig} illustrates the general behavior of the redshift
formula (\ref{zphi}) versus the azimuthal angle $\varphi $ for the
co-rotating massive test particles. This figure indicates how the frequency
shift in the Kerr background modifies with the motion of the geodesic
particle and how it changes for non-zero values of the rotation parameter.
The non-vanishing value of the redshift at the LOS ($\varphi =0$) for the
continuous green curve (referring to the standard Schwarzschild black hole)
quantifies the magnitude of the gravitational redshift. The difference between the other
curves and the green one is due to the dragging effect produced by the
rotation nature of the Kerr black hole. Furthermore, the total frequency
shift is maximal close to the midline ($\varphi \approx \pm \pi /2$), and
therefore, it is easier to be measured observationally.

On the other hand, one may note that in the Newtonian limit $\tilde{M},%
\tilde{a}\rightarrow 0$, the redshift formula of the Kerr black hole $%
z_{_{Kerr}}$ presented in Eq. (\ref{zphi}) reduces to

\begin{equation}
z_{Newton}=\tilde{M}^{1/2}\sin (\varphi +\delta )+\mathcal{O}\left( \tilde{M}%
\right),
\end{equation}%
which is the projection of the Keplerian velocity of a circularly orbiting
particle on the LOS, as it should be.

As the next step, we are going to express the mass-to-distance ratio $M/D$\
and spin-to-distance ratio $a/D$ of the Kerr black hole\ in terms of
directly measurable quantities for two special important cases of the
midline $\varphi \approx \pi /2$ and close to the LOS $\varphi \approx 0$.
These two cases have significant importance due to the fact that they
describe the frequency shift of emitters within accretion disks circularly
orbiting supermassive black holes hosted at the core of AGNs and there are
available astrophysical data from these specific points.

\subsection{Frequency shift on the midline}

When the position vector of orbiting objects with respect to the black hole
location is approximately orthogonal to the observer's LOS ($k^{r}=0$), we
can measure the high frequency shifted photons on the midline where $\varphi
=\pm \pi /2$. Therefore, for the highly redshifted ($\varphi =+\pi /2$) and
blueshifted ($\varphi =-\pi /2$) photons, the frequency shift formula (\ref%
{zphi}) reduces to 
\begin{equation}
1+z_{Kerr_{1,2}}^{(m)}\!\approx \frac{\left( 1-2\tilde{M}\right) +\tilde{M}%
^{1/2}\left( \tilde{a}\pm \tilde{\Delta}_{e}^{1/2}\right) }{\left( 1-2\tilde{%
M}\right) \sqrt{1-3\tilde{M}+2\,\tilde{a}\,\tilde{M}^{1/2}}},
\label{zKerrMid}
\end{equation}%
in the limit $\delta \rightarrow 0$\ that is the case for the real
astrophysical systems where the angular distance\ is of the order of
milliarcseconds. Here and in what follows, the index \textquotedblleft $m$"
means the observational quantities should be measured on the midline. This
relation is obtained for the co-rotating photon sources and the plus (minus)
sign refers to the redshifted (blueshifted) photons denoted by $%
z_{_{Kerr_{1}}}^{(m)}$ ($z_{_{Kerr_{2}}}^{(m)}$) on the midline. By
multiplying $R_{m}:=1+z_{Kerr_{1}}^{(m)}$\ and $B_{m}:=1+z_{Kerr_{2}}^{(m)}$%
, one can find the mass-to-distance ratio and spin-to-distance ratio of the
Kerr black hole in terms of directly observational elements on the midline
as follows \cite{pBaHmMuNprd2022}
\begin{equation}
\frac{M}{D}=\frac{R_{m}B_{m}-1}{2R_{m}B_{m}}\delta _{m}:=\mathcal{M}%
(R_{m},B_{m})\delta _{m},  \label{MDmid}
\end{equation}%
\begin{eqnarray}
&&\frac{a}{D}=\frac{\delta _{m}}{\left( 2R_{m}B_{m}\right) ^{3/2}\sqrt{%
R_{m}B_{m}-1}}\times   \notag \\
&&\left[ \left( R_{m}-B_{m}\right) ^{2}-\left( R_{m}+B_{m}\right) \sqrt{%
R_{m}^{2}+B_{m}^{2}-2R_{m}^{2}B_{m}^{2}}\right]   \notag \\
 && \ \ \   :=\mathcal{A}(R_{m},B_{m})\delta _{m},  \label{aDmid}
\end{eqnarray}%
up to the first order in $\delta _{m}$ in consistency with the condition $%
U_{d}^{\mu }=\delta _{t}^{\mu }$ for distant observers. In order to derive
the aforementioned relations, we also employed the approximation $%
r_{e}\approx D\delta _{m}$ given in Eq. (\ref{AppMid}) where the observable $%
\delta _{m}$ should be measured on the midline. Equations (\ref{MDmid})-(\ref%
{aDmid}) show that $M/D$ and $a/D$ of the Kerr black hole\ are expressed in
terms of directly measurable quantities $\left\{ R_{m},B_{m},\delta
_{m}\right\} $, but they are degenerate and we shall disentangle them in the
upcoming section.

\subsection{Frequency shift close to the line of sight}

Calculating the frequency shift of emitters in the vicinity of the LOS of
their orbital motion is also important for astrophysical applications.
Therefore, for angles close to zero on either side of the LOS where $\varphi
_{s}\rightarrow 0$ and $\delta _{s}\rightarrow 0$, one finds the frequency
shift formula (\ref{zphi}) reduces to%
\begin{eqnarray}
&&1+z_{Kerr_{1,2}}^{(s)}\approx   \label{zKerrphismall} \\
&&\frac{\left( 1-2\tilde{M}\right) +\tilde{M}^{\frac{1}{2}}\left[ \tilde{a}%
\pm \left( 1-2\tilde{M}\right) ^{-\frac{1}{2}}\tilde{\Delta}%
_{e}^{3/2}(\varphi _{s}+\delta _{s})\right] }{\left( 1-2\tilde{M}\right) 
\sqrt{1-3\tilde{M}+2\tilde{M}^{\frac{1}{2}}\tilde{a}}},  \notag
\end{eqnarray}%
for the slightly redshifted $z_{Kerr_{1}}^{(s)}$\ and slightly blueshifted $%
z_{Kerr_{2}}^{(s)}$\ photons. Here and in what follows, the index
\textquotedblleft $s$" corresponds to the measurements close to the LOS. $%
z_{Kerr_{1,2}}^{(s)}$ are obtained for the co-rotating particles only, and
the angles $\delta _{s}$\ and $\varphi _{s}$\ should be measured close to
the LOS. Note that $\delta _{s}$\ is an observable parameter whereas $%
\varphi _{s}$ is not, and they are related to each other through the
approximation (\ref{AppLOS}). The relations (\ref{zKerrphismall}) reduce to
the gravitational redshift $z_{g}=z_{Kerr_{1}}^{(s)}=z_{Kerr_{2}}^{(s)}$
exactly at the LOS where $\varphi _{s}=0=\delta _{s}$, and interestingly,
one can quantify the dragging effect at this point explicitly.

Unlike the midline case, one finds that $z_{Kerr_{1,2}}^{(s)}$\ are rather
complicated to find $M/D$ and $a/D$\ in terms of observables $\left\{
z_{Kerr_{1}}^{(s)},z_{Kerr_{2}}^{(s)},\delta _{s}\right\} $. On the other
hand, since $R_{m}$ and $B_{m}$ of the midline case are easier to be
measured observationally, it is important to incorporate them in the LOS
case where the \emph{redshift rapidity} is maximal (see the next section and
the right panel of Fig. \ref{zDotFig}). Therefore, by noting that $\tilde{M}%
=M/r_{e}$, we substitute $\tilde{M}=M/(D\delta _{m})=\mathcal{M}%
(R_{m},B_{m}) $\ from Eq. (\ref{MDmid})\ into (\ref{zKerrphismall}) and
solve $\left( 1+z_{Kerr_{1}}^{(s)}\right) +\left(
1+z_{Kerr_{1}}^{(s)}\right) $\ to obtain%
\begin{eqnarray}
&&\frac{a\varphi _{s}}{D\delta _{s}}=\frac{(1-2\mathcal{M})}{\sqrt{\mathcal{M%
}}}\left[ -1+\left( \frac{R_{s}+B_{s}}{2}\right) ^{2}\right. \times  \notag
\\
&&\left. \left( (1-2\mathcal{M})-\sqrt{(1-2\mathcal{M})^{2}-\frac{4(1-%
\mathcal{M})}{(R_{s}+B_{s})^{2}}}\right) \right]  \notag \\
&:&=\mathcal{F}(R_{m},B_{m},R_{s},B_{s}),  \label{aDlos}
\end{eqnarray}%
for the spin-to-distance ratio of the Kerr black hole close to the LOS that
is a function of a set of purely measurable elements $\left\{
R_{s},B_{s},R_{m},B_{m}\right\} $\ with $R_{s}:=1+z_{Kerr_{1}}^{(s)}$\ and $%
B_{s}:=1+z_{Kerr_{2}}^{(s)}$. To derive this relation, we applied the limits 
$\delta _{s},\varphi _{s}\rightarrow 0$ and employed the approximation $%
r_{e}\approx D\delta _{s}/\varphi _{s}$\ from Eq. (\ref{AppLOS}) valid close
to the LOS. One may note that the unobservable parameter $\varphi _{s}$\ can
be expressed in terms of the set of purely observational quantities $\left\{
R_{s},B_{s},\delta _{s};R_{m},B_{m},\delta _{m}\right\} $ by substituting $%
a/D$\ from (\ref{aDmid})\ into (\ref{aDlos}).

\section{\emph{Redshift rapidity} in Kerr background}

\label{SecRedshiftRapidity}

The mass-to-distance ratio (\ref{MDmid}) in the static limit $a\rightarrow 0$
reduces to the corresponding one in the standard Schwarzschild spacetime. So
far this relation has been employed to estimate the mass-to-distance ratio\
of several supermassive Schwarzschild black holes hosted at the center of galaxies \cite%
{ApJL,TXS,TenAGNs,FiveAGNs,TwoAGNs}. However, note that the mass-to-distance
ratio (\ref{MDmid}) as well as spin-to-distance ratios (\ref{aDmid}) and (%
\ref{aDlos}) are degenerate. In order to disentangle the mass $M$\ and spin $%
a$\ of the Kerr black hole as well as its distance from the Earth $D$ and
express them through closed formulae in terms of observational quantities,
we employ the \emph{redshift rapidity}, a new concept introduced in the
standard Schwarzschild spacetime recently in \cite{mMpBaHuNepjc2024}. In a
similar manner to \cite{mMpBaHuNepjc2024}, one can define the \emph{redshift
rapidity} in the Kerr spacetime background at the emission point as the
proper time evolution of the frequency shift $z_{Kerr}$ (\ref{zcircorbits})
in the following way%
\begin{equation}
\left. \dot{z}_{Kerr}\right\vert _{r=r_{e}}=\frac{dz_{Kerr}}{d\tau }=\frac{d%
}{d\tau }(U_{e}^{t}-b_{\varphi }U_{e}^{\varphi }),  \label{RR0}
\end{equation}%
where $\tau $\ is the proper time. Since the observer measures the \emph{%
redshift rapidity} here on the Earth, we rewrite Eq. (\ref{RR0}) at the
detector position as 
\begin{equation}
\dot{z}_{Kerr}=\frac{dz_{Kerr}}{dt}=\frac{d\tau }{dt}\frac{dz_{Kerr}}{d\tau }%
=\frac{1}{U_{e}^{t}}\frac{d}{d\tau }(U_{e}^{t}-b_{\varphi }U_{e}^{\varphi }),
\label{RR}
\end{equation}%
in which we used $U_{e}^{t}=\left. \frac{dt}{d\tau }\right\vert _{r=r_{e}}$%
,\ and $\dot{z}_{Kerr}$ is an observable quantity now. For the photon
sources circularly orbiting the Kerr black hole in the equatorial plane, $%
U_{e}^{t}$\ (\ref{tCompVelo}) and $U_{e}^{\varphi }$\ (\ref{phiCompVelo}) of
the $4$-velocity are constant, whereas the light bending parameter (\ref%
{bVsPhiDelta}) is time-dependent through $\delta $\ and $\varphi $.
Therefore, the \emph{redshift rapidity} relation (\ref{RR}) reduces to 
\begin{equation}
\dot{z}_{Kerr}=-\frac{db_{\varphi }}{d\tau }\ \frac{U_{e}^{\varphi }}{%
U_{e}^{t}}=-\left( \frac{\partial b_{\varphi }}{\partial \varphi }+\frac{%
\partial b_{\varphi }}{\partial \delta }\frac{\partial \delta }{\partial
\varphi }\right) \frac{\left( U_{e}^{\varphi }\right) ^{2}}{U_{e}^{t}},
\end{equation}%
where we used $U_{e}^{\varphi }=\left. \frac{d\varphi }{d\tau }\right\vert
_{r=r_{e}}$ in the last step and note that $\delta $\ is related to $\varphi 
$\ through Eq. (\ref{deltaPhiRel}).

By considering the light bending parameter (\ref{bVsPhiDelta}) and the $%
\delta $ function (\ref{deltaPhiRel}), one can perform the aforementioned
derivatives\ to find the \emph{redshift rapidity} in the Kerr background for
an arbitrary point on the circular motion as follows 
\begin{eqnarray}
&&\dot{z}_{Kerr} =\frac{DMr_{e}^{4}\text{$\Delta $}_{e}^{3/2}(D-r_{e}\cos
\varphi )\cos (\varphi +\text{$\delta $})}{\left[ r_{e}^{4}\left( 1-\frac{2M%
}{r_{e}}\right) \cos ^{2}(\varphi +\text{$\delta $})+\text{$\Delta $}%
_{e}^{2}\sin ^{2}(\varphi +\text{$\delta $})\right] ^{\frac{3}{2}}}  \notag
\\
&&\times \frac{\left( D^{2}+r_{e}^{2}-2Dr_{e}\cos \varphi \right) ^{-1}}{%
\left( r_{e}^{2}+a\sqrt{Mr_{e}}\right) \sqrt{r_{e}^{2}+2a\left(
Mr_{e}\right) ^{\frac{1}{2}}-3Mr_{e}}},  \label{RedRap}
\end{eqnarray}%
which reduces to the \emph{redshift rapidity} in the Schwarzschild spacetime
background in the limit $a\rightarrow 0$\ \cite{mMpBaHuNepjc2024}.

The profile of the \emph{redshift rapidity} versus the azimuthal angle $%
\varphi $ in the Kerr background is illustrated in Fig. \ref{zDotFig} for various values of the rotation parameter. This
figure shows that $\dot{z}_{Kerr}$ is maximal at the LOS, and therefore it
is easier to be measured for small angles $\varphi \approx 0$. One can also
see that the \emph{redshift rapidity} increases as the rotation parameter
increases.

Interestingly, in the Newtonian limit $M/r_{e}\rightarrow 0$ and $%
a/r_{e}\rightarrow 0$, the \emph{redshift rapidity} formula (\ref{RedRap})
reduces to the projection of the Keplerian acceleration of a circularly
orbiting massive test particle on the LOS 
\begin{equation}
\dot{z}_{Newton}=\frac{M}{r_{e}^{2}}\cos (\varphi +\delta )+\mathcal{O}%
\left( \frac{M^{2}}{r_{e}^{3}}\right) ,
\end{equation}%
for large distances $r_{e}/D\rightarrow 0$, as we expected.

Now, one can use the \emph{redshift rapidity} formula in Kerr spacetime (\ref%
{RedRap}) as well as $M/D$ and $a/D$ given in Eqs. (\ref{MDmid})-(\ref{aDmid}%
) on the midline to disentangle the Kerr black hole mass $M$, its rotation
parameter $a$,\ and its distance from Earth $D$, thus obtaining analytic
relations for these parameters at these specific points. On the other hand, we
can employ Eq. (\ref{RedRap}) and $a/D$ on the LOS given in Eq. (\ref{aDlos}%
) in order to find an analytic formula for the distance to Kerr black holes
in terms of the frequency shift on the midline and the \emph{redshift
rapidity} close to the LOS.

In what follows, we analyze two important cases of the midline and close to
the LOS describing water vapor clouds within accretion disks circularly
orbiting supermassive black holes hosted at the center of AGNs for which
data is available for frequency shift and \emph{redshift rapidity}.

\subsection{\emph{Redshift rapidity} on the midline}

The emitted photons on the midline are highly redshifted/blueshifted whereas
they have low \emph{redshift rapidity}. The mass-to-distance ratio and
spin-to-distance ratio of the central Kerr black hole are presented through
Eqs. (\ref{MDmid})-(\ref{aDmid}) in terms of observational
redshifted/blueshifted detected photons and the angular distance of their
source. On the other hand, by setting $\varphi =\varphi _{m}=\pi /2$ as well
as introducing the approximations (\ref{AppMid}) and $\delta _{m}\rightarrow
0$ in the \emph{redshift rapidity} formula (\ref{RedRap}), one can obtain $%
\dot{z}_{Kerr}^{(m)}$ corresponding to the same redshifted/blueshifted
photons on the midline as follows%
\begin{eqnarray}
&&\dot{z}_{Kerr}^{(m)}\approx \frac{\frac{M}{D\delta _{m}}}{D\left( \frac{a}{%
D\delta _{m}}\sqrt{\frac{M}{D\delta _{m}}}+1\right) }\times
\label{RedRapMid} \\
&&\frac{1}{\left[ \left( \frac{a}{D\delta _{m}}\right) ^{2}-2\frac{M}{%
D\delta _{m}}+1\right] ^{\frac{3}{2}}\sqrt{2\frac{a}{D\delta _{m}}\left( 
\frac{M}{D\delta _{m}}\right) ^{\frac{1}{2}}-3\frac{M}{D\delta _{m}}+1}}. 
\notag
\end{eqnarray}

As the final step, we solve Eqs. (\ref{MDmid})-(\ref{aDmid}) and (\ref%
{RedRapMid}) to find the mass of the Kerr black hole, its rotation
parameter, and its distance from the Earth as below%
\begin{eqnarray}
M=\frac{\mathcal{M}^{2}\mathcal{X}}{\dot{z}_{Kerr}^{(m)}\left( \mathcal{A}%
\sqrt{\mathcal{M}}+1\right) }\delta _{m},
\label{Mass}
\end{eqnarray}%
\begin{eqnarray}
a=\frac{\mathcal{AMX}}{\dot{z}_{Kerr}^{(m)}\left( \mathcal{A}\sqrt{\mathcal{M%
}}+1\right) }\delta _{m},
\label{Rotation}
\end{eqnarray}%
\begin{eqnarray}
D\!\!=\frac{\mathcal{MX}}{\dot{z}_{Kerr}^{(m)}\left( \mathcal{A}\sqrt{%
\mathcal{M}}+1\right) },
\label{Distance}
\end{eqnarray}%
\begin{eqnarray}
\mathcal{X}=\frac{1}{\sqrt{2\mathcal{AM}^{1/2}-3\mathcal{M}+1}\left( 
\mathcal{A}^{2}-2\mathcal{M}+1\right) ^{3/2}},
\label{Xparameter}
\end{eqnarray}%
in terms of the set of fully observational quantities $\left\{
z_{Kerr_{1}}^{(m)},z_{Kerr_{2}}^{(m)},\dot{z}_{Kerr}^{(m)},\delta
_{m}\right\} $\ that should be measured on the midline. One can also obtain
the radius of the emitter, which is not an observational element, with the
help of Eqs. (\ref{Distance})\ and (\ref{AppMid})\ as%
\begin{eqnarray}
r_{e}=\frac{\mathcal{MX}}{\dot{z}_{Kerr}^{(m)}\left( \mathcal{A}\sqrt{%
\mathcal{M}}+1\right) }\delta _{m}.
\label{emitterRadius}
\end{eqnarray}

Now, it would be helpful to find an analytic relation for the distance $D$\
in terms of observations close to the LOS since the \emph{redshift rapidity}
is maximal at this point and easier to be measured. In addition, it would be
nice to have this distance in terms of redshift and blueshift on the midline
as well where these quantities are maximal. We shall obtain such a distance
to the Kerr black hole in the next subsection.

\subsection{\emph{Redshift rapidity} close to the line of sight}

Here we take into account the slightly frequency-shifted emitters close to
the LOS with maximal \emph{redshift rapidity} and the spin-to-distance ratio
of the corresponding Kerr black hole is presented in Eq. (\ref{aDlos}) in
terms of the redshift/blueshift for $\varphi _{s}\approx 0$. First, from the
approximation (\ref{AppLOS}), we see that $r_{e}=aD\delta _{s}/\left(
a\varphi _{s}\right) $ where $a\varphi _{s}/\left( D\delta _{s}\right) $ is
given in terms of observations in Eq. (\ref{aDlos}). Hence, we replace $r_{e}
$\ with the aforementioned relation in the \emph{redshift rapidity} formula (%
\ref{RedRap}). Next, by keeping the term $a\varphi _{s}/\left( D\delta
_{s}\right) $, we divide the rest of $M$\ and $a$\ parameters by $D\delta
_{m}$\ to get 
\begin{widetext}
\begin{eqnarray}
\dot{z}_{Kerr}^{(s)} &\approx &\frac{\left( \frac{a\varphi _{s}}{D\delta _{s}%
}\right) ^{3}\left( \frac{M}{D\delta _{m}}\right) \left[ \left( \frac{%
a\varphi _{s}}{D\delta _{s}}\right) ^{2}\left( \frac{a}{D\delta _{m}}\right)
-2\left( \frac{a\varphi _{s}}{D\delta _{s}}\right) \left( \frac{M}{D\delta
_{m}}\right) +\frac{a}{D\delta _{m}}\right] ^{3/2}}{D\left( \frac{a}{D\delta
_{m}}\right) \text{$\delta $}_{m}\left[ \frac{a\varphi _{s}}{D\delta _{s}}%
-\left( \frac{a}{D\delta _{m}}\right) \text{$\delta $}_{m}\right] \left[
\left( \frac{a\varphi _{s}}{D\delta _{s}}\right) ^{3/2}\sqrt{\frac{M}{%
D\delta _{m}}}+\sqrt{\frac{a}{D\delta _{m}}}\right] \left[ \frac{a}{D\delta
_{m}}-2\left( \frac{a\varphi _{s}}{D\delta _{s}}\right) \left( \frac{M}{%
D\delta _{m}}\right) \right] ^{3/2}}\times   \notag \\
&&\frac{1}{\sqrt{2 \left( 
\frac{a\varphi _{s}}{D\delta _{s}}\right) ^{3/2}\left( \frac{a}{D\delta _{m}}%
\right) ^{1/2}\left( \frac{M}{D\delta _{m}}\right) ^{1/2}-3\left( \frac{%
a\varphi _{s}}{D\delta _{s}}\right) \left( \frac{M}{D\delta _{m}}\right) +%
\frac{a}{D\delta _{m}}}},  \label{RedRapLOS}
\end{eqnarray}%
\end{widetext}for the \emph{redshift rapidity} of photon sources close to
the LOS and we applied the limits $\delta _{s},\varphi _{s}\rightarrow 0$ as
well. Finally, we substitute Eqs. (\ref{MDmid})-(\ref{aDmid}) and (\ref%
{aDlos}) into (\ref{RedRapLOS}) and solve the resultant relation for $D$ to
obtain the distance as%
\begin{eqnarray}
&&D=\frac{\mathcal{F}^{3}\mathcal{M}\left( \mathcal{AF}^{2}+\mathcal{A}-2%
\mathcal{FM}\right) ^{3/2}}{\dot{z}_{Kerr}^{(s)}\mathcal{A}\text{$\delta _{m}
$}\sqrt{\mathcal{A}-3\mathcal{FM}+2\mathcal{F}\left( \mathcal{AFM}\right)
^{1/2}}}\times   \notag \\
&&\frac{1}{(\mathcal{F-A}\text{$\delta _{m}$})(\mathcal{A}-2\mathcal{FM}%
)^{3/2}\left( \sqrt{\mathcal{A}}+\mathcal{F}\sqrt{\mathcal{FM}}\right) },
\label{DistanceLOS}
\end{eqnarray}%
fully in terms of the set of observational quantities $\left\{
z_{Kerr_{1}}^{(s)},z_{Kerr_{2}}^{(s)},\dot{z}%
_{Kerr}^{(s)};z_{Kerr_{1}}^{(m)},z_{Kerr_{2}}^{(m)},\delta _{m}\right\} $\
where $\left\{ z_{Kerr_{1}}^{(s)},z_{Kerr_{2}}^{(s)},\dot{z}%
_{Kerr}^{(s)}\right\} $\ should be measured close to the LOS and $\left\{
z_{Kerr_{1}}^{(m)},z_{Kerr_{2}}^{(m)},\delta _{m}\right\} $\ should be
measured on the midline. From an observational point of view, this is a
significant formula for the Kerr black hole distance from the Earth since it
is a combination of highly redshifted\ photons on the midline and maximum 
\emph{redshift rapidity} on the LOS, and therefore, they are easier to be
measured. 

As the final remark, we would like to stress that the analytic formulae for
the Kerr black hole mass $M$ (\ref{Mass}), its rotation parameter $a$\ (\ref%
{Rotation}), and its distance from the Earth $D$\ (\ref{Distance}) \& (\ref%
{DistanceLOS}) are presented fully in terms of directly measurable
quantities of the accretion disk, and they are among the crucial findings of
the present study. In addition, the complete forms of the frequency shift
relation (\ref{zphi}) and \emph{redshift rapidity} formula (\ref{RedRap})
can be employed in black hole parameter estimation studies for an arbitrary
point on the circular motion of the emitter in order to obtain $M$, $a$, and 
$D$ of Kerr black holes. For instance, the general expressions (\ref{zphi})
and (\ref{RedRap}) can find astrophysical applications for modeling the
accretion disks consisting of\ water masers revolving supermassive black
holes hosted at the core of AGNs (see \cite%
{ApJL,TXS,TenAGNs,FiveAGNs,TwoAGNs}\ for the Schwarzschild black hole modelings).


\section{Discussion and final remarks}

\label{Discussion}

In this paper, we have considered the frequency shift of photon sources
circularly orbiting a Kerr black hole in the equatorial plane for an arbitrary point on the circular motion of the orbiting particles by taking into account the non-zero angular distance. Then for a special case on the midline, we represented
the mass-to-distance ratio and spin-to-distance ratio of the Kerr black hole
only in terms of observable quantities, that have been already obtained in \cite{pBaHmMuNprd2022}. Besides, we have
calculated the spin-to-distance ratio of the Kerr black hole close to the LOS
in terms of observational elements. On the other hand, we have computed the
proper time evolution of the frequency shift in order to calculate the \emph{%
redshift rapidity} in the Kerr spacetime background. Then, we employed the 
\emph{redshift rapidity} on the midline to express the Kerr black hole mass
and spin as well as its distance from the Earth fully in terms of directly
measurable quantities. In this scenario, the \emph{redshift rapidity} is a
general relativistic invariant observational parameter as well that is
calculated as the derivative of the frequency shift with respect to the
proper time in the Kerr black hole spacetime. In addition, we employed the 
\emph{redshift rapidity} close to the LOS as well as the mass-to-distance
ratio and spin-to-distance ratio on the midline in order to obtain the
distance to the Kerr black hole. This distance formula is significantly
important from an observational point of view since it is a function of
highly redshifted photons on the midline and maximum\emph{\ redshift rapidity%
} on the LOS, i.e., the quantities that are easier to be measured.

We have derived concise and elegant analytic formulae on the midline and
close to the LOS that allow us to disentangle the mass and spin of the Kerr
black hole as well as its distance from a distant observer, hence being able
to estimate these parameters separately. Our analytic formulae are valid on
the midline and close to the line of sight, having direct application to
supermassive black holes hosted at the core of active galactic nuclei
orbited by water vapor clouds within their accretion disks. The generic
exact expressions are valid for an arbitrary point of the emitter's orbit,
and they can be employed in black hole parameter estimation studies. 

Indeed, one can apply the generic exact relations to real astrophysical
systems where water maser clouds are circularly orbiting a central
supermassive black hole in the equatorial plane \cite{EdgeOn2011,Moran,MCP1}%
, as was accomplished in the Schwarzschild black hole case \cite%
{ApJL,TXS,TenAGNs,FiveAGNs,TwoAGNs}. As a concrete example, we refer to
supermassive black holes hosted at the core of the galaxies NGC 4258 and NGC
2273 where their accretion disks made of water masers circularly orbiting
the central object in the equatorial plane, and such megamaser systems have
been tracked for several galaxies using astrometry techniques. On the other
hand, the analytic formulae on the midline and close to the LOS can be
applied to the photon sources located at these special positions where can
be detected for most megamaser systems (see \cite{EdgeOn2011,Moran} for
instance).


\section*{Acknowledgements}

The author is grateful to A. Herrera-Aguilar and U. Nucamendi for carefully
reading the manuscript. He also acknowledges SNII and was supported by the
National Council of Humanities, Sciences, and Technologies of Mexico
(CONAHCyT) through Estancias Posdoctorales por M\'{e}xico Convocatoria
2023(1) under the postdoctoral Grant No. 1242413.

\appendix*

\section{($\protect\varphi +\protect\delta $)-dependent light bending
parameter in Kerr spacetime}

\label{bphi}

\begin{figure*}[t]
\centering
\includegraphics[scale=.9]{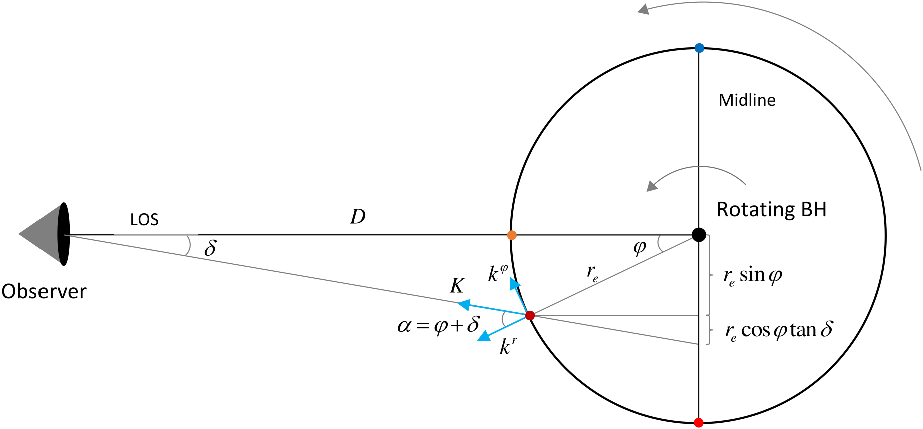}
\caption{The schematic illustration of a rotating black hole described by the line element (\ref{metric}) and a co-rotating photon source far enough from the black hole. This figure also shows the propagation direction $K$ of
the photons, and the relation between the unobservable azimuthal angle $%
\protect\varphi $ and the observable angular distance $\protect\delta $.}
\label{PhiDeltaFig}
\end{figure*}

In this appendix, we combine section V of \cite{pBaHmMuNprd2022}\ and the
appendix of \cite{mMpBaHuNepjc2024}\ to obtain the light bending parameter
of the photon sources for an arbitrary point on their circular orbit in the
equatorial plane mainly for the self-consistency of the article. Indeed, the
angular distance $\delta $\ has been neglected in \cite{pBaHmMuNprd2022} due
to the fact that the black holes are located far from the Earth. But even
though $\delta $ is small, its dependency on time is important in the
derivation of the \emph{redshift rapidity} formula, hence we take into
account its contribution to the light bending parameter in this article.

Fig. \ref{PhiDeltaFig} shows that the frequency-shifted photons can arrive
at the detector from any generic point on their circular motion in the
equatorial plane. The light bending parameter of such generic photons in the
axially symmetric spacetime backgrounds of the form (\ref{metric}) can be
calculated from the equation of motion of null particles ($g_{\mu \nu
}k^{\mu }k^{\nu }=0$) laying in the equatorial plane as%
\begin{equation}
g_{tt}\left( k^{t}\right) ^{2}+g_{rr}\left( k^{r}\right) ^{2}+2g_{t\varphi
}k^{t}k^{\varphi }+g_{\varphi \varphi }\left( k^{\varphi }\right) ^{2}=0,
\label{PhotonEOM}
\end{equation}%
where $k^{t}$\ and $k^{\varphi }$\ can be written in terms of the constants
of motions $E_{\gamma }$ and $L_{\gamma }$ with the help of the temporal
Killing vector field $\xi ^{\mu }=\delta _{t}^{\mu }$\ and the rotational
Killing vector field $\psi ^{\mu }=\delta _{\varphi }^{\mu }$ as follows 
\begin{equation}
k^{t}=\frac{E_{\gamma }g_{\varphi \varphi }+L_{\gamma }g_{t\varphi }}{%
g_{t\varphi }^{2}-g_{tt}g_{\varphi \varphi }},  \label{tCompOf4momentum}
\end{equation}%
\begin{equation}
k^{\varphi }=-\frac{E_{\gamma }g_{t\varphi }+L_{\gamma }g_{tt}}{g_{t\varphi
}^{2}-g_{tt}g_{\varphi \varphi }}.  \label{PhiCompOf4momentum}
\end{equation}

By substituting Eqs. (\ref{tCompOf4momentum})-(\ref{PhiCompOf4momentum})
into the equation of motion (\ref{PhotonEOM}), \ one finds $k^{r}$ in terms
of the metric components and constants of motion in the following form \ 
\begin{equation}
\left( k^{r}\right) ^{2}=\frac{g_{tt}L_{\gamma }^{2}+2g_{t\varphi }L_{\gamma
}E_{\gamma }+g_{\varphi \varphi }E_{\gamma }^{2}}{g_{rr}\left( g_{t\varphi
}^{2}-g_{tt}g_{\varphi \varphi }\right) }.  \label{rCompOf4momentum}
\end{equation}

As illustrated in Fig. \ref{PhiDeltaFig}, we define the auxiliary
bidimensional vector $K$ with the decomposition%
\begin{equation}
k^{r}=K\cos \alpha ,  \label{kr}
\end{equation}%
\begin{equation}
r_{e}k^{\varphi }=K\sin \alpha ,  \label{kphi}
\end{equation}%
with%
\begin{equation}
K^{2}=\left( k^{r}\right) ^{2}+r_{e}^{2}\left( k^{\varphi }\right) ^{2}.
\label{Kdef}
\end{equation}%
where the angular distance is non-zero and the angle $\alpha =\varphi
+\delta $ specifies the propagation direction $K$\ of the photons and the
radial component of the $4$-wave vector $k^{r}$ at the emission point. By
substituting Eqs. (\ref{PhiCompOf4momentum})-(\ref{rCompOf4momentum})\ into (%
\ref{Kdef}), one finds 
\begin{equation}
K^{2}=\frac{g_{tt}L_{\gamma }^{2}+2g_{t\varphi }L_{\gamma }E_{\gamma
}+g_{\varphi \varphi }E_{\gamma }^{2}}{g_{rr}\left( g_{t\varphi
}^{2}-g_{tt}g_{\varphi \varphi }\right) }+r^{2}\frac{\left( E_{\gamma
}g_{t\varphi }+L_{\gamma }g_{tt}\right) ^{2}}{\left( g_{t\varphi
}^{2}-g_{tt}g_{\varphi \varphi }\right) ^{2}},
\end{equation}%
in terms of the metric components and constants of motion. In addition, one
can find a similar relation for $K^{2}$ by replacing Eq. (\ref{kr})\ into (%
\ref{rCompOf4momentum}) as follows%
\begin{equation}
K^{2}=\frac{g_{tt}L_{\gamma }^{2}+2g_{t\varphi }L_{\gamma }E_{\gamma
}+g_{\varphi \varphi }E_{\gamma }^{2}}{g_{rr}\left( g_{t\varphi
}^{2}-g_{tt}g_{\varphi \varphi }\right) \cos ^{2}\alpha }.
\end{equation}

By equating the afformentioned equations for $K^{2}$, we obtain the
following relation for $b_{\varphi }$%
\begin{eqnarray}
&&\left( g_{tt}b_{\varphi }^{2}+2g_{t\varphi }b_{\varphi }+g_{\varphi
\varphi }\right) \left( g_{t\varphi }^{2}-g_{tt}g_{\varphi \varphi }\right)
\sin ^{2}\alpha   \notag \\
&&-r^{2}g_{rr}\left( g_{tt}b_{\varphi }+g_{t\varphi }\right) ^{2}\cos
^{2}\alpha =0,
\end{eqnarray}%
which gives the following solution for the $\alpha $-dependent light bending
parameter 
\begin{eqnarray}
&&b_{\varphi }=-\frac{g_{t\varphi }}{g_{tt}}-\frac{\left( g_{t\varphi
}^{2}-g_{tt}g_{\varphi \varphi }\right) \sin (\varphi +\delta )}{g_{tt}}%
\times   \label{bGamma} \\
&&\frac{1}{\sqrt{\left( g_{t\varphi }^{2}-g_{tt}g_{\varphi \varphi }\right)
\sin ^{2}(\varphi +\delta )-r_{e}^{2}g_{tt}g_{rr}\cos ^{2}(\varphi +\delta )}%
},  \notag
\end{eqnarray}%
that describes the light bending for a generic point of the circular orbit
on the equatorial plane in the axially symmetric black hole spacetimes of
the form (\ref{metric}).

On the other hand, we note that the angular distance $\delta $ is observable
whereas the azimuthal angle $\varphi $ is unobservable. In order to get rid
of the unobservable parameter $\varphi $, we consider far-away emitters such
that the geometrical configuration presented in Fig. \ref{PhiDeltaFig} holds
and the approximation $\alpha \approx \varphi +\delta $ is valid. This
assumption is necessary to break the degeneracy of the mass-to-distance\
ratio and spin-to-distance ratio in the Kerr black hole spacetime. This
allows us to extract a relation between $\delta $ and $\varphi $ as follows
(see Fig. \ref{PhiDeltaFig})%
\begin{equation}
D\sin \delta =r_{e}\sin \left( \varphi +\delta \right) ,  \label{PhiDeltaGR}
\end{equation}%
where $\delta $\ is observable only and $D$\ is the radial distance between the black hole and the detector. Since it is not possible to locate the
black hole's position on the sky with the help of observations, the rest of
the parameters $\left\{ D,r_{e},\varphi \right\} $\ are unknown. For the
distant observers where $D\gg r_{e}$\ and $\delta \rightarrow 0$, one finds
that this relation reduces to 
\begin{equation}
r_{e}\approx D\delta _{m},\text{ \ for \ }\varphi _{m}=\pm \frac{\pi }{2},
\label{AppMid}
\end{equation}%
\begin{equation}
r_{e}\approx \frac{D\delta _{s}}{\varphi _{s}},\text{ \ for \ }\varphi
_{s}\rightarrow 0,  \label{AppLOS}
\end{equation}%
for the midline $\varphi =\varphi _{m}=\pm \pi /2$\ and close to the LOS $%
\varphi =\varphi _{s}\rightarrow 0$, respectively. One can solve the
relation (\ref{PhiDeltaGR})\ and express $\delta $\ in terms of the rest of
the parameters as follows%
\begin{equation}
\delta =\arccos \left( \frac{D-r_{e}\cos \varphi }{\sqrt{%
D^{2}+r_{e}^{2}-2Dr_{e}\cos \varphi }}\right),  \label{deltaPhiRel}
\end{equation}%
and note that we chose the acceptable solution of the angular distance since as 
$\varphi $\ decreases/increases, $\delta $\ should decrease/increase in the
same direction as well (see Fig. \ref{PhiDeltaFig}).


\end{document}